\documentclass[english]{article}
\usepackage[T1]{fontenc}
\usepackage[latin9]{inputenc}
\usepackage{babel}
\usepackage{mathrsfs}
\usepackage{url}
\usepackage{graphicx}
\usepackage[unicode=true]
 {hyperref}
\usepackage{breakurl}
\usepackage{lineno}

\makeatletter
\newcommand{\lyxaddress}[1]{
\par {\raggedright #1
\vspace{1.4em}
\noindent\par}
}

\makeatother

\begin{document}

\title{\textbf{Plasma Wakefield Linear Colliders - Opportunities and Challenges}}

\author{Erik Adli \\ Department of Physics, University of Oslo, N-0316 Oslo, Norway}

\maketitle

\lyxaddress{\begin{center}
\par\end{center}}
\begin{abstract}
\thispagestyle{empty}A linear electron-positron collider operating
at TeV scale energies will provide high precision measurements and
allow, for example, precision studies of the Higgs boson as well as
searches for physics beyond the standard model. A future linear collider
should produce collisions at high energy, with high luminosity and
with a good wall plug to beam power transfer efficiency. The luminosity
per power consumed is a key metric that can be used to compare linear
collider concepts. The plasma wakefield accelerator has demonstrated
high-gradient, high-efficiency acceleration of an electron beam, and
is therefore a promising technology for a future linear collider.
We will go through the opportunities of using plasma wakefield acceleration
technology for a collider, as well as a few of the collider-specific
challenges that must be addressed in order for a high-energy, high
luminosity-per-power plasma wakefield collider to become a reality.
\\
\end{abstract}
\newpage{}

\pagenumbering{arabic}\setcounter{page}{1}

\section*{Introduction}

A high-energy, high-luminosity electron-positron linear collider,
providing clean collisions between fundamental particles, will provide
measurements complementing the LHC results. Such a machine
will increase our understanding of the TeV-scale, and be sensitive
to beyond-Standard Model physics above the LHC energies. Two international
projects have proposed a linear collider based on RF technology: the
International Linear Collider, ILC \cite{ILC_TDR}, and the Compact
Linear Collider, CLIC \cite{CLIC_CDR}.  The main linac of the ILC
will use superconducting RF cavities operating at an accelerating
gradient of 31.5 MV/m. ILC will have a footprint (length)
of 20-50 km with centre-of-mass energies ranging from 250-1000 GeV.
The main linac of CLIC will use normal conducting X-band RF structures
operating at an accelerating gradient of 100 MV/m. The machine will
have a footprint of 11-50 km, with centre-of-mass energies from 380
- 3000 GeV. The Future Circular Collider study, FCC \cite{FCC_CDR} proposes a circular electron-positron collider with centre-of-mass energy up to 365 GeV. The designs for the linear colliders have been on-going for several
decades in order to develop technology and optimize choices and machine parameters.
Despite a rich physics program, the current global funding climate
makes the realization of the above colliders challenging.
A linear collider based on RF technology with centre-of-mass energies in
the say 10 TeV range seem even more unlikely to be realized due
to cost and footprint constraints.

Beam-driven plasma wakefield acceleration (PWFA) may be an alternative
route to TeV-scale e- e+ collisions, possibly enabling a higher energy
reach. To be attractive, new concepts should show improvement of some
form with respect to the existing projects (ILC, CLIC). The very high
gradients demonstrated by PWFA give promise of reducing, possibly
drastically, the footprint of a collider, or increase the energy for
the same footprint. Electron acceleration with gradients of 10s of GV/m has been demonstrated
in PWFA experiments \cite{EA:FFTB_E-1}, and the acceleration of electron
beams and positron beams with multi-GeV gradients has also been achieved \cite{EA:FACET_E-1,EA:FACET_P}. In comparison,
CLIC will operate at gradients of up to 100 MV/m. By assuming an average
gradient of 1 GV/m for a future plasma-based collider \cite{EA:ANA_LC-1},
beams of up to 3.5 TeV could be produced in the same main linacs 3.5
km tunnels of a CLIC machine at 380 GeV \cite{EA:CLIC380}, assuming
the beam delivery system is upgraded to handle the energy increase.
As very high accelerating gradients are well established for PWFA,
this paper will focus on other aspects of a collider. High luminosity
is as important for precision physics studies as high centre-of-mass
energy. Preliminary physics studies show that the luminosity requirements
set for ILC and CLIC remain as high - or higher - at higher collision
energies \cite{EA:ROLOFF2018}. For linear colliders a key performance
metric is the luminosity per power. Taking into account beam strahlung \cite{EA:BEAMSTR},
the useful luminosity is optimized when beams are flat, and the luminosity
per power, $\mathscr{L}/P_{AC}$, scales as \cite{CLIC_CDR}

\noindent 
\begin{equation}
\mathscr{L}/P_{AC}\propto\eta_{\mathrm{AC\rightarrow beam}}\frac{1}{\sqrt{\sigma_{z}}\sqrt{\beta_{y}\varepsilon_{y}}}.\label{eq:EA_lum}
\end{equation}

To maximize luminosity, the vertical emittance $\varepsilon_{^{y}}$
and the vertical focusing function $\beta_{y}$ must be minimized, the
wall-plug-to-beam efficiency $\eta_{\mathrm{AC\rightarrow beam}}$
must be maximized, and bunches of short length $\sigma_{z}$ must
be collided. For bunch lengths much shorter than those of CLIC, beamstrahlung considerations \cite{EA:BEAMSTR} must be taken into account,
and the useable luminosity may not scale as Eq. (\ref{eq:EA_lum}).
Thus, in addition to reduction of footprint, one should also ask,
based on Eq. (\ref{eq:EA_lum}): how could plasma wakefield acceleration
improve the luminosity per power? Of course, the ultimate metric for
comparing collider proposals would be luminosity per cost,
however, cost comparisons would require an advanced stage of collider
design. Cost will not be further discussed in this paper. Currently there are no solutions for the acceleration of collider quality positron beams in plasmas. While PWFA positron acceleration is an area of active research, progress towards an electron positron collider is likely to be limited until a solution for positron acceleration of collider beams has been established.

\section*{Existing collider concepts}

In order to discuss the opportunities and challenges of a PWFA-based
linear collider (PWFA-LC) with respect to existing technology it is
useful to go through some key aspects of linear collider projects,
in particular the CLIC two-beam acceleration scheme due to its parallels
to beam-driven plasma wakefield collider concepts. In both cases,
energy extracted from drive beams are used to accelerate the main
beams to be collided. In a PWFA-LC, the drive beam energy is extracted by
the plasma, and transferred from the plasma to a trailing, co-linear
main beam.  In CLIC, X-band structures extract the RF energy and
transfer it to RF accelerating structures in a parallel main beam line.
CLIC uses the two-beam scheme to efficiently and robustly generate
short X-band RF pulses. The short pulses allow operation of normal
conducting accelerating structures at 100 MV/m \cite{CLIC_CDR}. The
efficiency of the two beam acceleration will be discussed below.

A good design based on any technology would require a global optimization
of the main machine parameters taking into account all parts of the
collider, much like has been done for RF collider design, i.e. CLIC \cite{CLIC_CDR}.
For plasma-based technologies, the understanding and description
of parameter dependencies is at present insufficient for performing design work towards a plasma-based collider. 

\begin{figure}
\begin{centering}
\includegraphics[width=1\columnwidth]{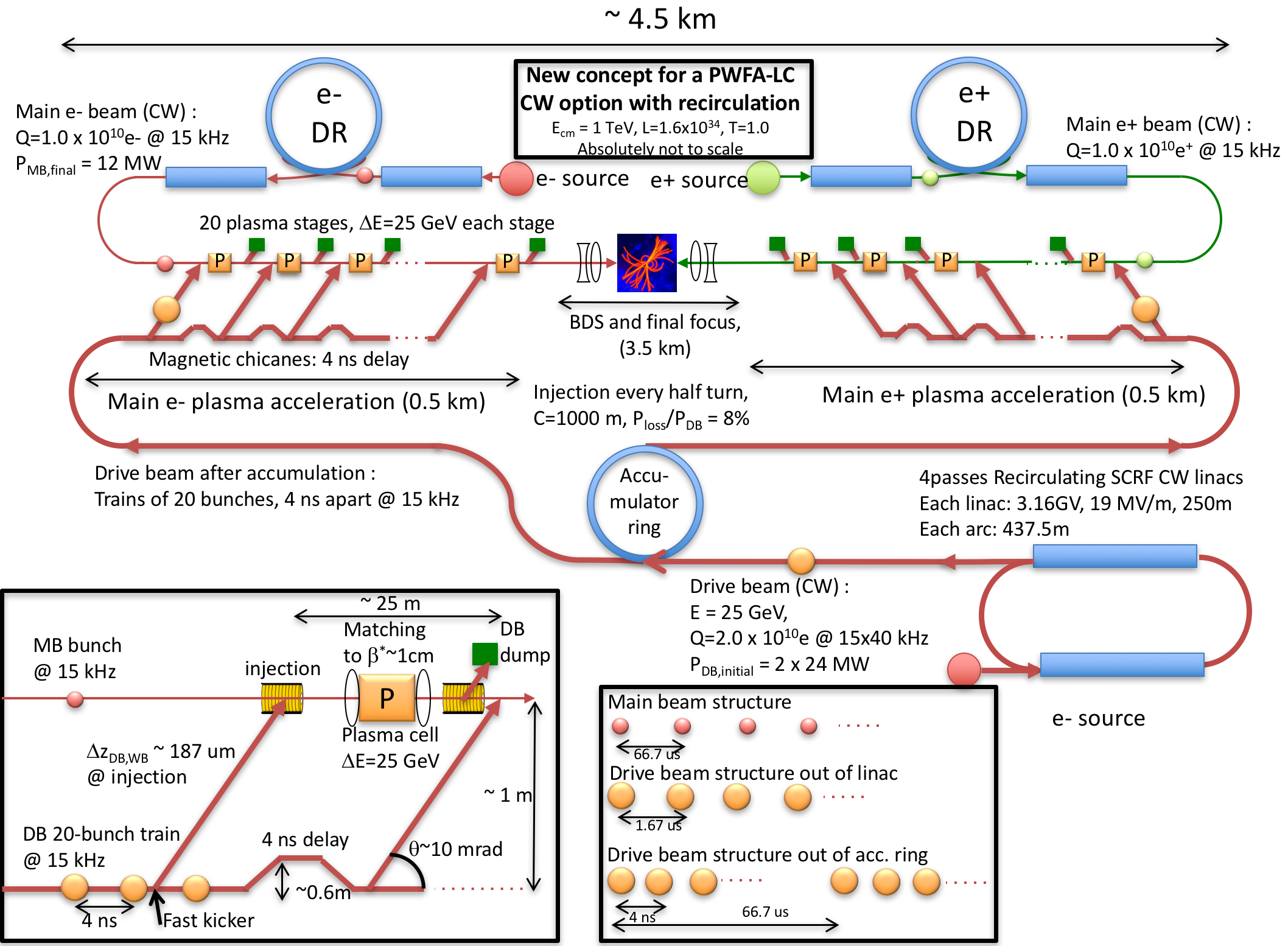}
\par\end{centering}
\emph{\caption{\label{fig:1}A concept for a PWFA linear collider, as put together
 for the US particle physics 2013 Community Summer Study \cite{EA:PWFA_LC}.
While far from being at the detail level of a machine design, this concept has stimulated progress towards a collider in a number
of areas, discussed in this paper: power efficiency numbers, drive
beam generation, drive beam distribution, staging, plasma lens research,
transverse instabilities and transverse tolerances.}
}
\end{figure}

Instead studies or ideas have been put forward for
plasma colliders inspired or based on the parameter optimization done
for CLIC and ILC, with the aim of improving certain parts of the machine.
Examples are : adding an afterburner to an already built RF collider
\cite{EA:AFT1,EA:AFT2}; replacing injectors, possibly also damping
rings and bunch compressors by plasma-based injectors \cite{EA:PLASMA_CATHODE};
improve/shorten the beam delivery system and final focus \cite{EA:CHEN_PL,EA:LINDSTROM2016};
making beam dumps more compact, potentially with some energy recovery
\cite{EA:GUO_TODO}; or, replacing the main linac with advanced accelerator
technology. We will in this review focus on the latter, the main linac,
since this part has been studied the most, and is the most costly
component of the current collider design. Several concepts have been
put forward in the literature, including \cite{EA:ROS_TODO,EA:PWFA_LC_OLD,EA:PWFA_LC},
with the aim of identifying the main challenges of PWFA-LC and to
establish base parameters for studying these challenges. In order
to discuss opportunities and challenges of a PWFA-LC in more detail
we start by considering the latest iteration, written
up for the US particle physics 2013 Community Summer Study \cite{EA:PWFA_LC},
illustrated in Figure \ref{fig:1}. 

Since it is currently not worked out how to perform a global optimization
of machine parameters, including how to choose the plasma density for the PWFA stages,
the parameter choices for \cite{EA:PWFA_LC} were made as follows:
assume parameters similar to those of the ILC for the main beams,
in order to ensure similar beam delivery performance and luminosity;
scale the plasma stages to provide 25 GeV energy gain, in a few meters of plasma, in order to reach an average gradient over
the main linac of at least 1 GV/m (``effective gradient''). The
choice of 25 GeV per stage was a compromise between minimizing stages and
components (driving towards high energy gain per stage) and a need
for practical drive beam parameters (driving towards lower energy
gain per stage). The plasma density was minimized in order to mitigate
sources of emittance growth in the plasma (instabilities, scattering
and others). To optimize the drive beam parameters, e- drive bunches
and e- main bunches in the blow-out regime \cite{EA:BLOWOUT1, EA:BLOWOUT2} were assumed. In the blow-out regime a bubble is formed by plasma electrons blown outwards by the driver, gathering in a sheath around an evacuated area filled with only ions. The ions form a uniform density ion channel creating a focusing force that varies linearly with radius. This focusing force preserves emittance of electron beams  \cite{EA:BLOWOUT1, EA:BLOWOUT2} as long as the ion motion does not significantly affect the electron beam. 

Furthermore, gaussian bunches as well as a drive-beam to main-beam transformer ratio of one were assumed - a conservative choice made
so that the performance would not rely on advanced bunch manipulation.
With the main beam parameters, the transformer ratio, and the plasma
density set, the drive beam parameters were derived according to
the procedure described in \cite{EA:TZOUFRAS}. Since presently it is unclear
what the best mechanism to accelerate positrons in a plasma is, a
clearly stated assumption in \cite{EA:PWFA_LC} - necessary to be
able to discuss overall collider concepts and parameters - was that
positrons are accelerated with the same performance as electrons.
The time structure suggested in \cite{EA:PWFA_LC} is uniformly spaced
colliding beams with a repetition rate of 5-30 kHz, though a pulsed time structure may also
be envisaged.  While the overall concept in \cite{EA:PWFA_LC} has so far not been further developed, 
many aspects have since publication been scrutinized and discussed with
experts in the conventional accelerator community (Fermilab, CLIC, ILC),
which has led to a number of constructive comments, stimulating further
work and progress towards a PWFA-LC in a number of areas: power efficiency
numbers, drive beam generation, drive beam distribution, staging,
plasma lens research, transverse instabilities and transverse tolerances.
We now discuss the progress in the different areas in more detail.

\section*{Efficiency}

\begin{figure}
\begin{centering}
\includegraphics[width=0.8\columnwidth]{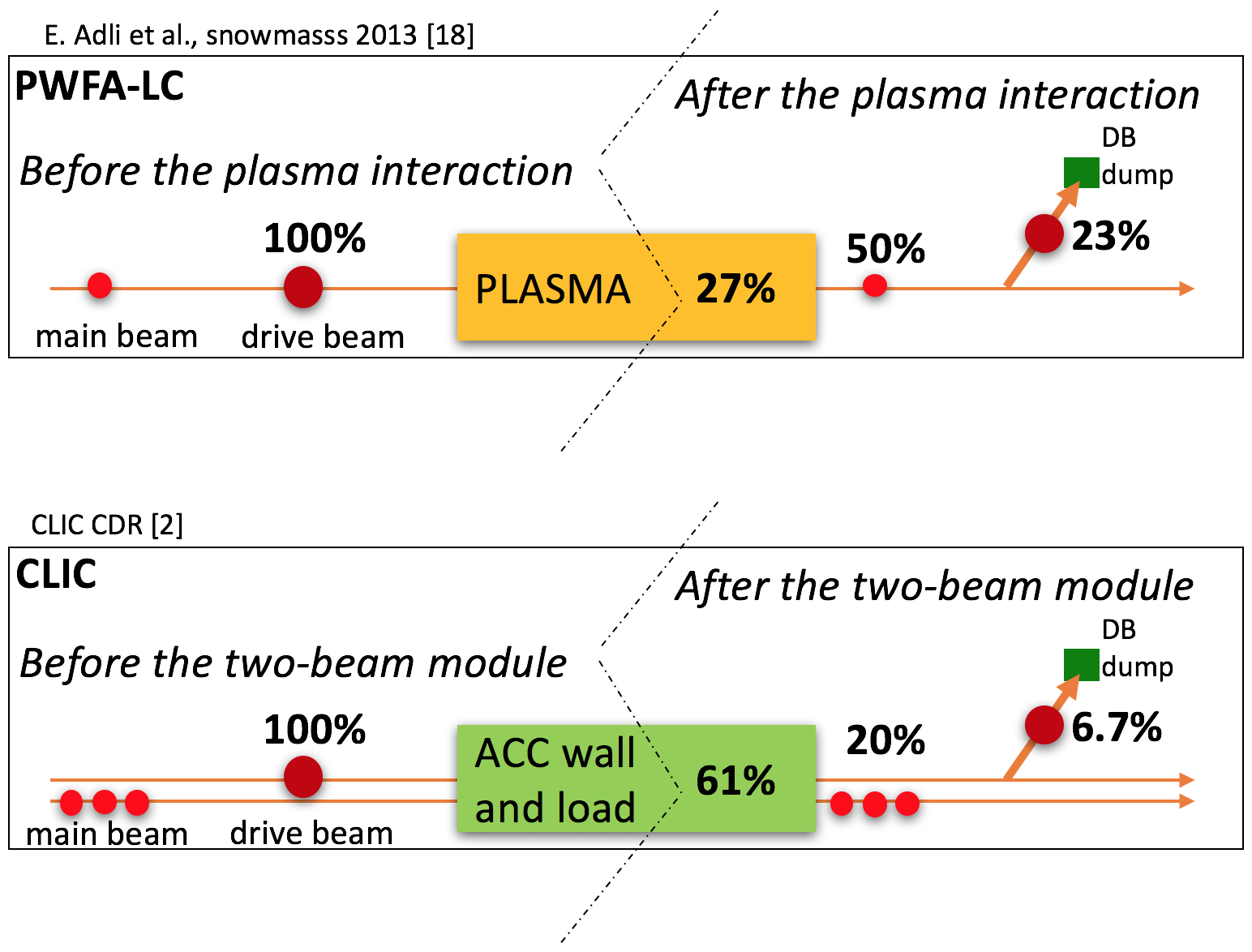}
\par\end{centering}
\emph{\caption{\label{fig:2}Drive beam to main beam efficiency, as reported for
the PWFA-LC in \cite{EA:PWFA_LC} (top) and as calculated for the CLIC
design \cite{CLIC_CDR} (bottom). The numbers indicate the fraction
of the energy of the bunches before and after acceleration. For the
PWFA-LC 50\% of the drive beam energy is transferred to the main beam, while 27\% of the energy is left in the plasma and
23\% in the drive beam going to the dump. The corresponding CLIC drive beam to main beam efficiency
is 20\%. An important distinction between the two estimates is that the
RF to beam efficiency of CLIC is constrained by transverse wakefields.
For the PWFA-LC, the efficiency estimates in \cite{EA:PWFA_LC} and in this figure transverse instabilities 
were not considered, since instabilities and their mitigation
are still a topic of research. Therefore the current PWFA-LC efficiency estimates may change once all constraints have been fully understood and taken into account.}
}
\end{figure}

In order to maximize the luminosity per power, the drive-beam to main-beam
(DB-to-MB) energy transfer efficiency has to be maximized. In PWFA
this efficiency has been shown in simulation to be very high; more
than 90\% is estimated in \cite{EA:TZOUFRAS} for optimally shaped
bunches, while 50\% is reported in \cite{EA:PWFA_LC} assuming Gaussian
bunches. More specifically, according to the simulations performed
in \cite{EA:PWFA_LC} the DB-to-wake efficiency is 77\% and the wake-to-MB
efficiency is 65\%, for a total of 77\%$\times$65\% = 50\% . The
first efficiency can possibly be increased by shaping the drive beam,
and the latter may possibly be increased by shaping the main beam,
by this approaching the higher numbers given in \cite{EA:TZOUFRAS}.
In PWFA two-beam experiments, wake-to-MB efficiencies of more than
30\% have been reported \cite{EA:FACET_E-1}, while proposed FACET-II
experiments aims at demonstrating 50\% DB-to-MB efficiency \cite{EA:FII}.
In comparison, corresponding numbers for the CLIC design are a DB-to-wake
efficiency of 81\% and a wake-to-MB efficiency of 25\%, resulting
in a total DB-to-MB efficiency of 20\% \cite{CLIC_CDR}. While the
CLIC efficiency is significantly lower than the PWFA-LC number, the
CLIC efficiency is constrained by transverse wakefields, limiting
the bunch charge and beam loading. In CLIC, a train with order of hundred bunches extracts the RF from a single fill of an accelerating cavity, while in \cite{EA:PWFA_LC} a single bunch extracts the energy from the plasma wake. The existing PWFA-LC studies have so far not considered the effect of transverse wakefields on the efficiency. An improved estimate of the efficiency for a PWFA-LC will require an improved understanding of the transverse instabilities and their mitigation mechanisms, the topic of the next section.  Figure
\ref{fig:2} summarizes the DB-to-MB efficiency estimates for the
the PWFA-LC and for CLIC.

\section*{Transverse instabilities}

While no attempt at quantifying transverse tolerances is done in \cite{EA:PWFA_LC},
theoretical descriptions of the transverse hosing instability in PWFA
exist \cite{EA:HOS1,EA:HOS2,EA:HOS3}, and simplified models of the
transverse instabilities have recently been suggested \cite{EA:PWFA_LIM1_FERMILAB-1,EA:PWFA_LIM1_FERMILAB-2,EA:LIM3}.
The simplified models aim at modeling the PWFA-instability in the
same language for describing the well-known BBU-instability in RF
accelerators; the transverse forces are expressed as a wake function,
parametrized only as a function of the plasma cavity size. This allows for
simple scaling laws and clear parameter dependencies required for
global optimization and thus improved designs for a PWFA-LC. While
the models put forward need to be further benchmarked with 3D simulations
and eventually experiments, the scalings - the transverse wakefield
increase as the inverse fourth power of the aperture - indicate that
the very small aperture of a plasma cavity compared to e.g. CLIC structures
lead to transverse wakes many orders of magnitude stronger in a PWFA-LC
than in CLIC (7-8 orders of magnitude stronger, assuming a CLIC aperture of about 1 mm and a typical
PWFA blow-out of a few 10 $\mu$m). In a PWFA-LC, the drive
beam defines the center of the plasma channel, and thus cannot be
offset with respect to this channel. However, inevitable transverse
jitter between the drive beam and the main beam implies that the main
beam intra-bunch wake may put significant constraints on the charge
\cite{EA:PWFA_LIM1_FERMILAB-2} and thus of PWFA-LC parameters and
performance. Such constraints have already been identified and addressed
in the design of normal conducting RF linacs, including the main linacs
of CLIC \cite{CLIC_CDR}, as discussed earlier. When a sufficiently
good understanding of the transverse instabilities and their mitigation
mechanisms have been obtained for PWFA, similar constraints should
be taken into account in a global parameter optimization. A number
of methods have been suggested for mitigating the PWFA transverse
instabilities. As discussed in \cite{EA:HOS2,EA:HOS3} they may be
grouped into three; reduction of instability seed \cite{EA:RAMP};
disruption of the coherence and reduction of the beam-plasma coupling. Disruption of the
coherence can be done in a variety of ways. The most straight forward
way is perhaps to induce a correlated energy spread in the beam in the form of BNS-damping
\cite{EA:BNS}, a technique well known in RF accelerators. BNS-damping was successfully applied to the Stanford Linear Collider \cite{EA:BNS_SLC} and will be used to stabilize the beam in the CLIC main linacs. BNS-damping will likely have similar effects for PWFA accelerators, and has been proposed in \cite{EA:HOS3,EA:PWFA_LIM1_FERMILAB-2}. The damping does
not have to rely on correlated energy spread; any effect that produces
a focusing varying along the beam may have a stabilizing effect if
the variation is tuned correctly. A recent innovative example of such
an effect is ion motion; while in itself a potential detrimental effect
for emittance preservation \cite{EA:ION1}, the attraction of ions
into the beam effectively produces a variable focusing with potential
for greatly mitigating the instability \cite{EA:ION2,EA:ION3,EA:ION4}.
However, whether the resulting emittance growth due to the ion motion
itself is acceptable needs to be studied further. To summarize, there
has recently been good progress both in the modeling of the BBU-instability
of PWFA and in the study of possible mitigation mechanisms. More detailed
studies, using collider parameters, are still needed to assess whether
the beam quality required for high luminosity collision can be achieved.
While transverse instabilities may lead to very tight transverse tolerances in order to prevent unacceptable emittance growth, 
centroid kicks resulting from the strong plasma focusing channels for beams
offset with respect to the channel center may lead to even tighter
tolerances. These kicks may lead to the two
colliding beams partly or fully missing each other, thus leading to
luminosity loss even if the beam emittances for each beam separately
may not significantly dilute. In \cite{EA:PWFA_LIM2_SCUHLTE-1} it
is shown that for the PWFA-LC parameters in \cite{EA:PWFA_LC}, the
alignment tolerances of the main beam is on the order of a few nm,
leading to very tight stability requirements for the drive and main
beams injected into the plasma cells, possibly tighter than those
arising from the transverse instabilities.

\begin{figure}
\begin{centering}
\includegraphics[width=0.8\columnwidth]{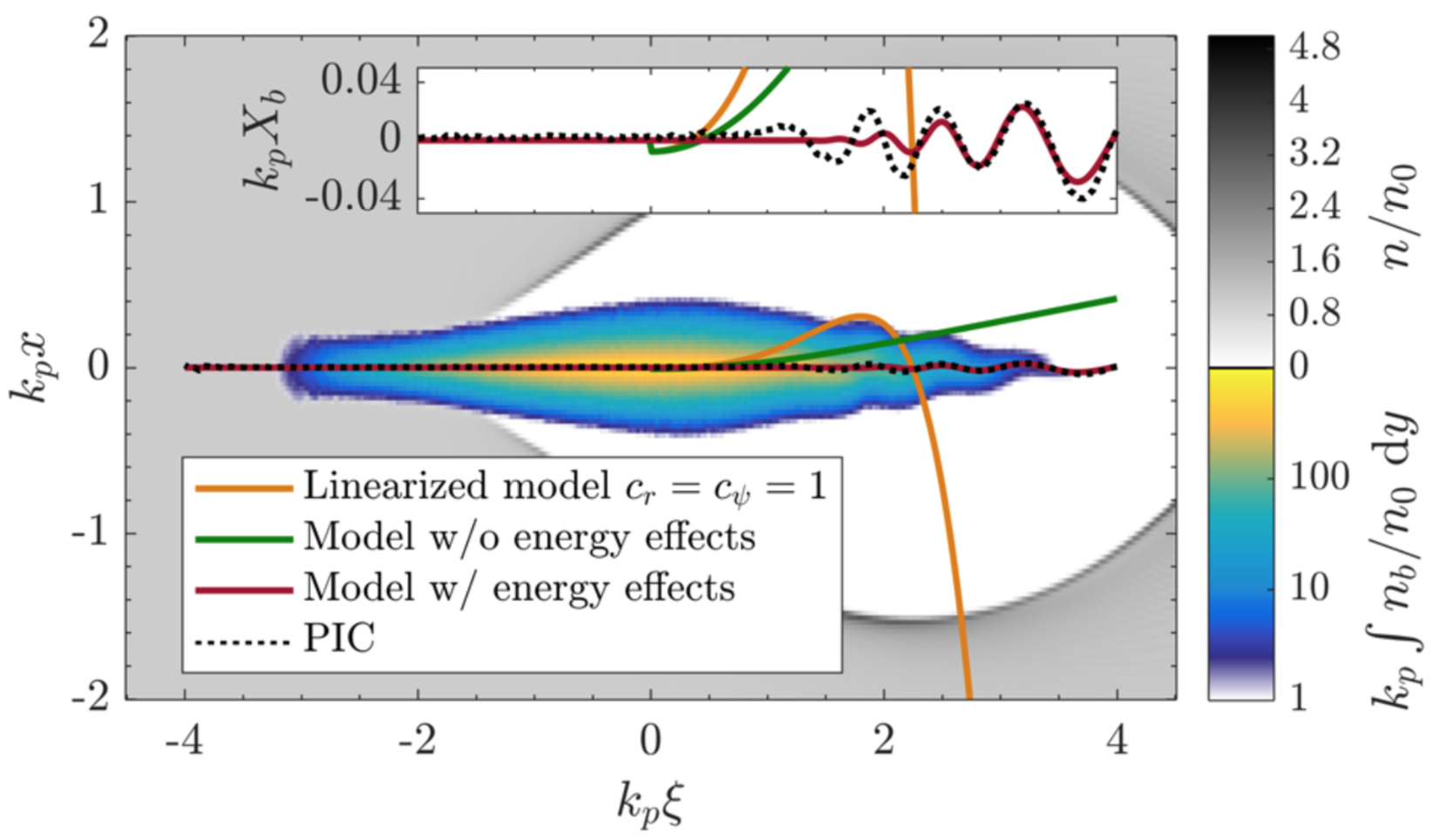}
\par\end{centering}
\emph{\caption{\label{fig:3} While transverse instabilities in PWFA and their mitigation
is still a subject of research, models show increasingly good correspondence
with full particle-in-cell (PIC) simulations. This figure from \cite{EA:HOS3}
shows the result of PIC simulations (dashed line) compared to mathematical
models of increasing complexity (solid lines) of a beam with an offset
with respect to the plasma sheaths, and where a correlated energy
spread mitigates the growth of the transverse instability. }
}
\end{figure}

\section*{Drive beam generation}

For a PWFA-LC scheme based on multiple plasma stages short, high-charge,
high-energy drive beams must be produced in an energy efficient manner.
In \cite{EA:PWFA_LC} it was proposed to use a superconducting RF recirculating
linac to produce the drive beam, and an accumulator ring for timely
distribution of the drive bunches. Superconducting RF is a mature technology,
and a wall-plug to drive beam efficiency estimate of 60\% or more
seems reasonable \cite{EA:PWFA_LIM2_SCUHLTE-1}. Refs. \cite{EA:PWFA_LIM2_SCUHLTE-1, EA:PWFA_IMP_3-1}
point out that the synchrotron energy loss in the accumulator ring
will lead to a too large bunch-to-bunch energy difference, for large
number of drive bunches, and proposed instead a linac that provides
the desired drive bunch time structure directly (equal spacing between
drive bunches) without the need for an accumulator ring. Such a linac would also be based on superconducting RF, using e.g. 1 GHz ILC-like cavities,
with a high efficiency. Although no further drive beam generation
design efforts have been reported after \cite{EA:PWFA_IMP_3-1,EA:PWFA_LIM2_SCUHLTE-1},
it is therefore likely that the drive beam generation can be significantly
simpler than indicated by the concept presented in \cite{EA:PWFA_LC}. 

\section*{Drive beam distribution}

\begin{figure}
\begin{centering}
\includegraphics[width=0.9\columnwidth]{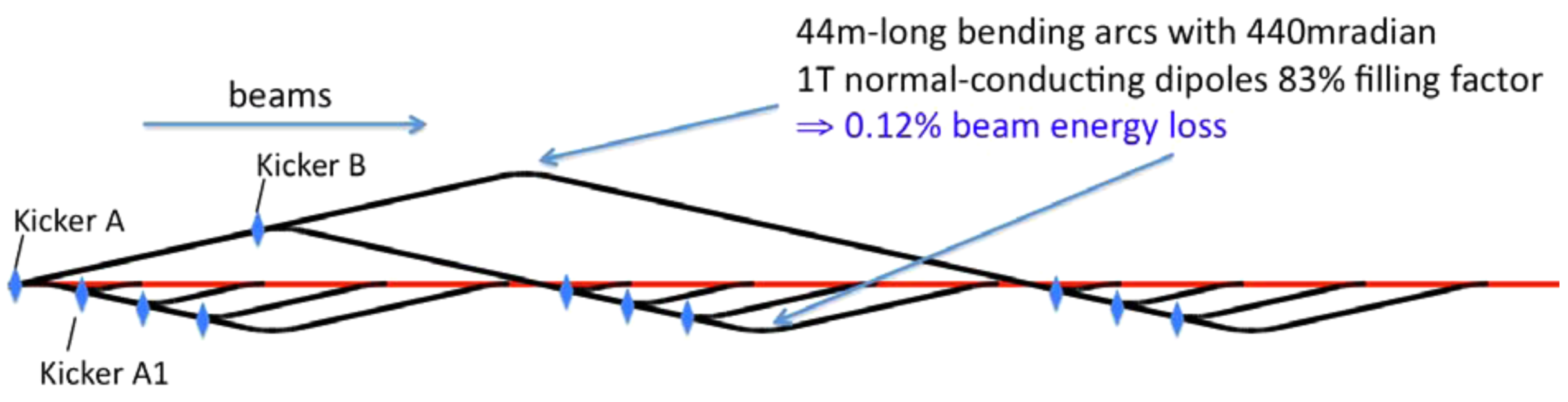}
\par\end{centering}
\emph{\caption{\label{fig:4}A tree-structure chicane concept for providing the individual
drive bunches with the appropriate delay in order to synchronize them
to the main beam, as suggested in \cite{EA:PWFA_IMP_3-1,EA:PWFA_LIM2_SCUHLTE-1}.
This concept reduces the total tunnel length, since separate bending
arcs are not needed for each drive bunch, while keeping the delay
path bending angles small.}
}
\end{figure}

Even if a drive beam train with the desired qualities has been generated and is correctly spaced, with e.g a few ns uniform spacing \cite{EA:PWFA_LC}, it is far from trivial to synchronize
individual drive bunches to the main beam, and injecting them on a
co-linear trajectory with the correct phase. In \cite{EA:PWFA_LC}
the idea was to use delay chicanes to synchronize the beams. At each
stage, the last bunch in the train would be ejected from the train
using fast kickers, and injected in front of the main beam, while
the rest of the train would be delayed by the chicane by the same
amount as the bunch spacing, see Figure \ref{fig:1}. For the drive beam energies assumed in \cite{EA:PWFA_LC}
bending magnets with fields of several Tesla would be required for
such chicanes, and as pointed out in \cite{EA:PWFA_IMP_3-1} this
may not be compatible with synchrotron radiation
losses in the strong bends. Having separate 180 degree bending arcs for each drive bunch
as suggested in \cite{EA:PWFA_LC_OLD} may be feasible, but may also become very
costly, based on ILC cost estimates. In \cite{EA:PWFA_IMP_3-1,EA:PWFA_LIM2_SCUHLTE-1}
an intermediate option, a tree-structure chicanes reducing the total
tunnel length while having reasonable bending angles, is proposed.
An example of such a tree-structure is shown in Fig \ref{fig:4}.
The drive beam is sent through separate tunnels that have an angle
with respect to the main linac. This angle is chosen to produce the
correct delay with respect to the main beam (a few degrees for the parameters in \cite{EA:PWFA_LC}). This concept, though likely not optimal, indicates
that ways to distribute high energy drive beams for a staged PWFA-LC
may be established. Also here, detailed
studies would be required to arrive at real designs and optimized
solutions.

\section*{Staging}

\begin{figure}
\begin{centering}
\includegraphics[width=0.8\columnwidth]{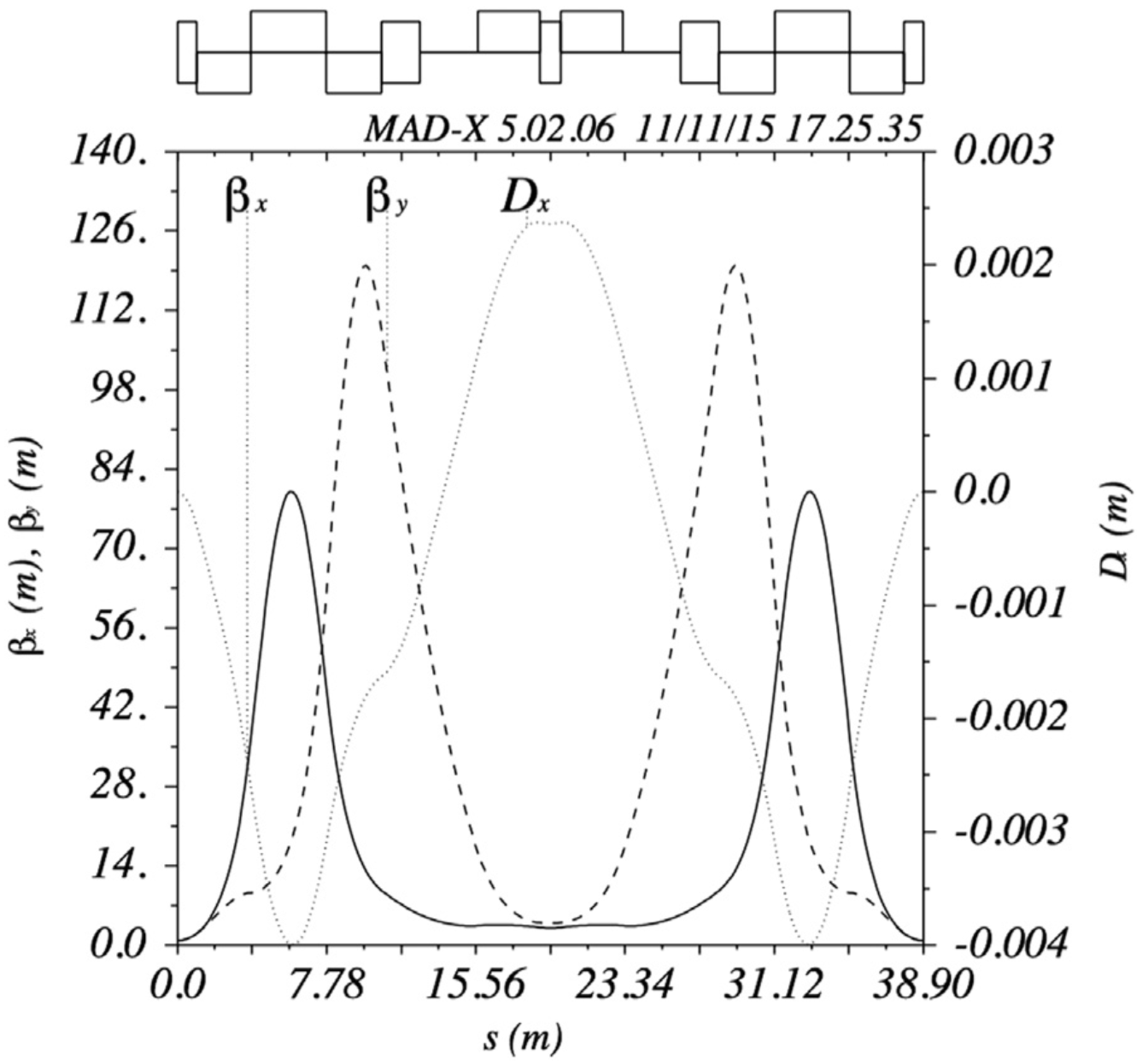}
\par\end{centering}
\emph{\caption{\label{fig:5}An example of achromatic optics for staging between
two plasma stages for 500 GeV beams, from \cite{EA:PWFA_LC_ISTAGE-1}.
The graph shows the beam focusing function $\beta$ in both planes,
as well as the dispersion function $D$ in the injection plane. Due
to the energy spread in the main beam, the chromatic errors as well
as dispersion arising from injection and extraction must be controlled
in order to avoid unacceptable emittance growth. }
}
\end{figure}

Since new drive beams need to be injected between plasma stages it
is foreseen that the main beam exits from plasma cells into vacuum,
before being reinserted into the next plasma cell. Tight transverse
and longitudinal tolerances on the drive beam and main beam must be
achieved in the interstage lattice, and in addition, main beam emittance
growth due to chromatic errors resulting from the very strong focusing
forces in the plasma cell must be contained. In \cite{EA:PWFA_LC}
no attempt were made to design an interstage lattice, however, two
studies have later been performed. Ref. \cite{EA:LINDSTROM2016} suggests
an elegant method for creating achromatic quadrupole-drift lattices,
while \cite{EA:PWFA_LC_ISTAGE-1} discusses various aspects that
need to be taken into account for interstage designs. By using the
matching methods from \cite{EA:LINDSTROM2016} a working example lattice
of 39 m for a 500 GeV beam line, fulfilling most requirements for
an example interstage design, is discussed in \cite{EA:PWFA_LC_ISTAGE-1}
and shown in Figure \ref{fig:5}. However, as it is pointed out in
\cite{EA:PWFA_LC_ISTAGE-1}, the length scaling of an interstage lattice
will necessarily increase as the square root of the beam energy, leading
to very long interstages for very high energy colliders. The exact
design of the interstage and the centre-of-mass energy of a future
collider will decide whether this is compatible with the assumption
of an 1 GV/m effective gradient. For the example in \cite{EA:PWFA_LC_ISTAGE-1},
a 1 GV/m effective gradient, or higher may be obtained for about 1
TeV of centre-of-mass energy, or lower.

\section*{Plasma lenses}

The recent years have seen good progress in the development of active plasma lenses \cite{EA:PL1}, an alternative to quadrupole magnet focusing of interest for addressing some of the staging challenges discussed above. In an active plasma lens, an axially symmetric focusing field is set up by a current pulse passing through a broken down gas confined inside a capillary \cite{EA:PL1}. The principle is similar to that of the Lithium lens \cite{EA:LIT}, however, with active plasma lenses having much less strong scattering and thus better potential for preserving collider quality beams. Selected recent results include better control of non-linearities \cite{EA:PL2, EA:PL3} and first demonstrations of emittance preservation for beams with $\mu$m-level normalized emittances \cite{EA:PL2, EA:PL3}.  At the same time, theoretical studies \cite{EA:PL4} indicate that the intensity of a collider beam may preclude the use of active plasma lenses in most part of a collider with parameters similar to those of \cite{EA:PWFA_LC}.

\section*{Positrons}

This review has been focused on the use of beam-driven plasmas for electron acceleration in the blow
out regime \cite{EA:BLOWOUT1, EA:BLOWOUT2}, which can not be used for positron acceleration.  We highlight again that at present it is unclear how to accelerate collider quality positron beams in a plasma, no matter which regime is used, see e.g.  \cite{EA:CORDE_POS}. Until new ideas for positron acceleration are conceived, we do not see a clear path towards a high-luminosity electron-positron collider based on PWFA.  However, an interesting alternative path to a Multi-TeV collider could
be to use two electron linacs to produce Multi-TeV photons through inverse Compton scattering instead; a gamma-gamma
collider \cite{EA:gg}. While the physics case for a $\gamma$-$\gamma$
collider as complement to e- e+ colliders has been investigated previously,
e.g. \cite{EA:gg_low}, physics performance studies for a $\gamma$-$\gamma$
collider as the only multi-TeV collider has started only recently \cite{EA:gg_high}. Preliminary findings \cite{EA:gg_high} indicate the physics potential is interesting \cite{EA:gg_high}, however, a multi-TeV $\gamma$-$\gamma$ should ideally be preceded by a first stage CLIC or ILC in order to access model independent measurements of the Higgs coupling through electron-positron collisions.

\section*{Summary}

It has been well established that both electrons and positrons can
be accelerated with gradients orders of magnitude higher than what
is done in RF cavities, giving promise of linear colliders with a
reduced footprint per centre-of-mass energy. In the recent years there
has been good progress in identifying challenges of a Multi-TeV plasma-based
collider, and the work to address them has started, i.e. for drive beam
generation, staging, transverse tolerances as discussed above. To go from current collider concepts to a real design, with a consistent
parameter set, will require a large amount of design work for all
sub-systems. A strong collaboration between the plasma accelerator
community and the RF linear collider communities, with their 30+ years
of experience, would be highly fruitful for this design effort, and to address numerous technical challenges (alignment power
flow, cooling, to name a few). Until there is an established working
regime for efficient positron acceleration with high beam quality,
a parameter set for a PWFA e- e+ collider cannot be established. An interesting
alternative could be to consider a Multi-TeV gamma-gamma collider.
The more mature PWFA-LC concepts become, the better experimental work towards a collider may be guided. Therefore it is likely
that more dedicated funding for design work (as opposed to funding
for experiments only) would help advance the progress towards a linear
collider based on plasma wakefield acceleration.

\section*{Acknowledgements}

The author would like to thank Kyrre N. Sjøbæk and Ben Chen for their good help with manuscript proofreading.

\end{document}